\documentclass[12pt]{iopart}
\usepackage{cite} 
\usepackage{iopams}
\usepackage{setstack}

\begin{document}

\title[Exactly solvable models for multiatomic molecular Bose-Einstein condensates]{Exactly solvable models for multiatomic molecular Bose-Einstein condensates}

\author{ G. Santos$^{1}$}

\address{$^{1}$ Instituto de F\'{\i}sica da UFRGS \\ 
Av. Bento Gon\c{c}alves, 9500 - Agronomia - Porto Alegre - RS - Brazil}
\ead{gfilho@if.ufrgs.br, gfilho@cbpf.br}

\begin{abstract}
I introduce two family of exactly solvable models for multiatomic hetero-nuclear and homo-nuclear molecular Bose-Einstein condensates
through the algebraic Bethe ansatz method. The conserved quantities of the respective models are also showed.
\end{abstract}

\section{Introduction}

One of the most interesting recent experimental achievements in physics is the one that led to realizations of Bose-Einstein condensates (BEC), by taking dilute alkali gases to extremely low temperatures \cite{early,angly}. Since then, a great effort has been devoted to the comprehension of  new phenomena involving this state of matter as well as its properties, either experimentally or theoretically. On the experimental side I could mention a molecular BEC compound that has been obtained 
combining different techniques \cite{mol}, leading this kind of  research also in the direction of a chemistry of BEC, 
where, for instance,  by Feshbach resonances \cite{Feshbach, RevFeshbach1, RevFeshbach2} or 
photo-association \cite{RevPhotoassociation1, RevPhoto2} the atomic constituents may form molecules. 

Many compounds of diatomic homo-nuclear molecular BECs \cite{HomoCs,HomoRb,HomoFeshbachCs2} have been produced since the first 
realization \cite{HomoFeshbachDonley}. Also, diatomic hetero-nuclear molecular BECs have been detected using these 
techniques \cite{HetphotoKRb1, HetPhotoKRb2, HetPhotoNaCs, HetPhotoRbCs, HetFeshbachRbRb, HetFeshbachKRb, HetFeshbachRbCs}. 
Actually, due to the rapid technological developments in the field of ultra-cold systems, it is believed that some of these 
experiments may be just the dawn of the study of multiatomic molecules \cite{HammerTeo,EsryEfimovCs}. More recently the 
experimental evidence for Efimov states in an ultra-cold cesium gas  \cite{EfimovEvidenceCs,Homo3BEfimovCs2,Homo4BEfimovCs} 
and a mixture of ultra-cold potassium and rubidium gases \cite{HetEfimovKRb} provides a physical ground for the investigation of triatomic and tetratomic homo-nuclear and triatomic hetero-nuclear molecular BECs.

These results boosted the search for solvable models that could describe some of the BEC properties \cite{jon1,jon2,jonjpa,key-3,dukelskyy,Ortiz,Kundu,eric5,GSantosaa,GSantos,GSantos1,GSantos2,GSantos3,Zhou1,Zhou2,Vardi,maj}. The rationale beneath these studies is that through exactly solvable models it  is possible to fully take into account quantum fluctuations, going beyond the usual mean field approximations. Then, I expect that this approach may provide some impact in this area, as well as a contribution to the field of integrable systems itself \cite{hertier, batchelor}. In this paper I will use the algebraic Bethe ansatz method. The algebraic formulation of the Bethe ansatz, and the associated quantum inverse scattering method (QISM), was primarily  developed by the group of mathematical physicists in St. Petersburg \cite{fst,ks,takhtajan,korepin,faddeev}. The QISM could be used to study the one-dimensional spin chains, quantum field theory in one-dimensional bosons interacting systems \cite{korepin1} and two-dimensional lattice models \cite{korepin2}, systems of strongly correlated electrons \cite{ek,ek2}, conformal field theory \cite{blz}, as well as precipitated the notion of quantum algebras (deformations of universal enveloping algebras of Lie algebras) \cite{jimbo85,jimbo86,drinfeld,frt}.  For a pedagogical and historical review see \cite{faddeev2}. 

Owing to recent insights in the understanding of the construction of Lax operators it is possible to obtain solvable models suitable for the effective description of the interconversion interactions occurring in the BEC.
Inspired by some of these ideas I present, in the present paper, the construction of two complete family of Bethe-ansatz solvable models 
for both homo-nuclear and hetero-nuclear  molecular BECs obtained through a combination of three Lax 
operators constructed using special realizations of the  $su(2)$ Lie algebra and of the Heisenberg-Weyl Lie algebra, as well as a multibosonic representation of the $sl(2)$ Lie algebra, discussed recently in \cite{Tomasz}. Notice that the models obtained through this construction do not have spatial degrees of freedom.

The paper is organized as follows: In Section 2, I will review shortly the algebraic Bethe ansatz method and present the Lax operators and the transfer matrix for both models. In Section 3, I will present a family of multiatomic homo-nuclear models and their solutions. In Section 4, I will present a family of multiatomic hetero-nuclear models and their solutions. In Section 5, I will make my remarks.

\section{Algebraic Bethe ansatz method}

In this section we will shortly review the algebraic Bethe ansatz method and present the transfer matrix used to get the solution of the models \cite{jonjpa,Roditi}. We begin with the $gl(2)$-invariant $R$-matrix, depending on the spectral parameter $u$,

\begin{equation}
R(u)= \left( \begin{array}{cccc}
1 & 0 & 0 & 0\\
0 & b(u) & c(u) & 0\\
0 & c(u) & b(u) & 0\\
0 & 0 & 0 & 1\end{array}\right),\end{equation}

\noindent with $b(u)=u/(u+\eta)$ and $c(u)=\eta/(u+\eta)$. Above,
$\eta$ is an arbitrary parameter, to be chosen later. It is easy
to check that $R(u)$ satisfies the Yang-Baxter equation

\begin{equation}
R_{12}(u-v)R_{13}(u)R_{23}(v)=R_{23}(v)R_{13}(u)R_{12}(u-v). 
\end{equation}

\noindent Here $R_{jk}(u)$ denotes the matrix acting non-trivially
on the $j$-th and the $k$-th spaces and as the identity on the remaining
space.

Next we define the monodromy matrix  $T(u)$,

\begin{equation}
T(u)= \left( \begin{array}{cc}
 A(u) & B(u)\\
 C(u) & D(u)\end{array}\right),
\end{equation}

\noindent that satisfy the Yang-Baxter algebra,

\begin{equation}
R_{12}(u-v)T_{1}(u)T_{2}(v)=T_{2}(v)T_{1}(u)R_{12}(u-v).\label{RTT}
\end{equation}

\noindent In what follows we will choose different realizations for the monodromy matrix $\pi(T(u))=L(u)$  
to obtain solutions of two family of models for multiatomic hetero-nuclear and homo-nuclear molecular BECs.
In this construction, the Lax operators $L(u)$  have to satisfy the relation

\begin{equation}
R_{12}(u-v)L_{1}(u)L_{2}(v)=L_{2}(v)L_{1}(u)R_{12}(u-v).
\label{RLL}
\end{equation}
\noindent where we use the notation,
\begin{equation}
L_1 = L(u) \otimes I  \;\;\; \mbox{and} \;\;\; L_2 = I \otimes L(u).
\end{equation}

Then, defining the transfer matrix, as usual, through

\begin{equation}
t(u)= tr \;\pi (T(u)) = \pi(A(u)+D(u)),
\label{trTu}
\end{equation}
\noindent it follows from (\ref{RTT}) that the transfer matrix commutes for
different values of the spectral parameter; i. e.,

\begin{equation}
[t(u),t(v)]=0, \;\;\;\;\;\;\; \forall \;u,\;v.
\end{equation}
\noindent Consequently, the models derived from this transfer matrix will be integrable. Another consequence is that the coefficients $\mathcal{C}_k$ in the transfer matrix $t(u)$,

\begin{equation}
t(u) = \sum_{k} \mathcal{C}_k u^k,
\end{equation}
\noindent are conserved quantities or simply $c$-numbers, with

\begin{equation}
[\mathcal{C}_j,\mathcal{C}_k] = 0, \;\;\;\;\;\;\; \forall \;j,\;k.
\end{equation}

If the transfer matrix $t(u)$ is a polynomial function in $u$, with $k \geq 0$, it is easy to see that,

\begin{equation}
\mathcal{C}_0 = t(0) \;\;\; \mbox{and} \;\;\; \mathcal{C}_k = \frac{1}{k!}\left.\frac{d^kt(u)}{du^k}\right|_{u=0}. 
\end{equation}

We will use three solutions of the equation (\ref{RLL}): 

\begin{itemize}

\item[(i)] \textbf{The $L^S(u)$ Lax operator:}

\begin{equation}
L^{S}(u)=\frac{1}{u} \left( \begin{array}{cc}
u-\eta S^{z} & -\eta S^{+} \\
-\eta S^{-} & u+\eta S^{z} \end{array}\right),
\label{LS}
\end{equation}

\noindent in terms of the $su(2)$ Lie algebra with generators $S^{z}$
and $S^{\pm}$ subject to the commutation relations

\begin{equation}
[S^{z},S^{\pm}]=\pm S^{\pm},\;\;\;\;[S^{+},S^{-}]=2S^{z}.
\label{su2}
\end{equation}

\item[(ii)] \textbf{The $L^j(u)$ Lax operator:}

\begin{equation}
L^{j}(u) = \left( \begin{array}{cc}
u + \eta N_j & j \\
j^{\dagger} & \eta^{-1} \end{array}\right),
\label{Lj}
\end{equation}

\noindent in terms of the Heisenberg-Weyl Lie algebra with generators $N_j$, $j$, $j^{\dagger}$ 
and $I$, subject to the commutation relations

\begin{equation}
[N_j,j]= - j,\;\;\;\;[N_j,j^{\dagger}]= +j^{\dagger},\;\;\;\;[j,j^{\dagger}] = I\;\;\;and\;\;\;[I,\star]=0,
\label{Heis}
\end{equation}

\noindent where $\star$ means $N_j,\;j$ or $\;j^{\dagger}$.

\item[(iii)] \textbf{The  $L^{A}(u)$ Lax operator:}

\begin{equation}
L^{A}(u) = \left( \begin{array}{cc}
u+\frac{\eta}{2} A_{0} & \eta A_{-} \\
-\eta A_{+} & u-\frac{\eta}{2} A_{0} \end{array}\right),
\label{LA}
\end{equation}
\noindent in terms of the $sl(2)$ Lie algebra with generators $A_{0}$ and $A_{\pm}$, subject to the commutation relations

\begin{equation}
[A_{-},A_{+}]= A_{0},\;\;\;\;[A_{0},A_{\pm}] = \pm 2A_{\pm}.\label{sl2}
\end{equation}

\end{itemize}

Using the co-multiplication properties of the Lax operator and the $gl(2)$ invariance of the $R$-matrix, we can 
obtain different realizations for the monodromy matrix:

\begin{itemize}

\item[(i)] \textbf{Multiatomic homo-nuclear:} For the multiatomic homo-nuclear molecular BEC model we choose

\begin{equation}
\pi(T (u)) = L(u) = \eta^{-1} G L^j (u - \delta - \eta^{-1} )L^A (u + \omega),
\label{Lutrihom}
\end{equation}
\noindent with, $G = diag(-,+)$, from which we find the following transfer matrix

\begin{eqnarray}
t(u) &=& -\eta^{-1}(u + \omega + \frac{\eta}{2} A_{0})(u-\delta-\eta^{-1}+\eta N_{b}) \nonumber \\ 
&+& \eta^{-2}(u + \omega - \frac{\eta}{2} A_{0})+bA_{+}+b^{\dagger}A_{-},
\label{tutrihom}
\end{eqnarray}
\noindent with 
\begin{equation}
t(0)=\eta^{-1}\omega(\delta+2\eta^{-1})-\omega N_{b} + \frac{1}{2}(\delta-\eta N_{b})A_{0} +bA_{+}+b^{\dagger}A_{-},
\label{t0trihom}
\end{equation}
\noindent and, discarding $c$-number terms, the conserved quantities are,

\begin{eqnarray}
\mathcal{C}_0 &=& t(0), \label{cqmhn1} \\
\mathcal{C}_1 &=& \frac{1}{2} A_0 + N_b \label{cqmhn2}.
\end{eqnarray}

\item[(ii)] \textbf{Multiatomic hetero-nuclear:} For the multiatomic hetero-nuclear molecular BEC models we choose

\begin{equation}
\pi(T (u)) = L(u) = \eta^{-1} u^{-} G L^S (u^{-} )L^A (u^{+} ),
\label{Lutrihet}
\end{equation}
\noindent with $u^{\pm} = u \pm  \omega$, $G = diag(+,-)$, from which we find the following transfer matrix

\begin{eqnarray}
t(u) & = & u^{-} A_{0} - 2u^{+} S^{z} + \eta (S^{+}A_{+} + S^{-}A_{-}),
\label{tutrihet}
\end{eqnarray}
\noindent and the conserved quantities,

\begin{eqnarray}
\mathcal{C}_0 &=& t(0), \label{cqmhtn1} \\
\mathcal{C}_1 &=& A_0 - 2S^{z} \label{cqmhtn2}.
\end{eqnarray}

\end{itemize}

    In the next sections we will describe the models and its integrability by the algebraic Bethe ansatz method, using 
different realizations of the algebras (\ref{su2}), (\ref{Heis}) and (\ref{sl2}). The Hamiltonians are 
written in the Fock space using the standard notation.  We are considering the coupling 
parameters real, such that the Hamiltonians are Hermitian. In the diagonal part of the Hamiltonians,  the $U_j$ parameters 
describe the atom-atom, atom-molecule and molecule-molecule $S$-wave scatterings and the $\mu_j$ parameters are the 
externals potentials. The operators $N_j$ are the number operators of atoms or molecules. In the off diagonal 
part of the Hamiltonians the parameter $\Omega$ is the amplitude for interconversion of atoms and molecules.

%
%

\section{Multiatomic homo-nuclear molecular models}
     
In this section we present the integrability of a new family of Hamiltonians describing  multiatomic homo-nuclear 
molecular BEC. The Hamiltonians that describes the interconversion of homogeneous molecules labelled 
by $b$ with $l$ atoms of type $a$ are given by

\begin{eqnarray}
H &=& U_{a}N_a^2 + U_{b}N_b^2 + U_{ab}N_aN_b + \mu_aN_a + \mu_bN_b \nonumber \\ 
&+& \Omega((a^{\dag})^l \; b \; \alpha_{-}(N_a) + \alpha_{-}(N_a) \; b^{\dag} \; (a)^l ),
\label{Hamulthom}
\end{eqnarray}
\noindent where $\alpha_{-}(N_a)$ is a function of $N_a$ that controls the amplitude of 
interconversion $\Omega$. This indicates that the density of atoms $N_a$ has some influence in the
generation of a bound-state composed by $l$ identical atoms. The $l=3$ case was studied in \cite{GSantos}. The total number of particles $N = N_a + lN_b$ is a conserved quantity.

There is a multibosonic realization of the $sl(2)$ Lie algebra \cite{Tomasz}

\begin{equation}
A_{0}=\alpha_{0}(N), \;\;\; A_{-} = \alpha_{-}(N)a^l, \;\;\; A_{+} = (a^{\dagger})^l\alpha_{-}(N),
\label{multsl2}
\end{equation}
\noindent with

\begin{eqnarray}
 \alpha_{0}(N) &=& \frac{2}{l}(N - R) + \alpha_{0}(R),  \\
 \alpha_{-}(N) &=& \sqrt{\frac{N!}{(N + l)!}(\frac{1}{l}(N - R) + \alpha_{0}(R))(\frac{1}{l}(N - R) + 1)}, 
\label{multsl2b}
\end{eqnarray}
\noindent where $N = a^{\dagger}a$ and $l \in \mathbb{N}$. The operator $R$ is

\begin{equation}
\label{cases}
R = \cases{0 & for $\;\;l=1$,\\
\frac{l-1}{2} + \sum_{m=1}^{l-1}\frac{e^{-(2\pi m/l)N}}{e^{(2\pi m/l)} - 1} & for  $\;\;l>1$,\\}
\end{equation}
\noindent and acts on the states $\{|n\rangle\}$ as  $R|n\rangle=n\;mod\;l|n\rangle$. The 
function  $\alpha_{0}(R)$ is a positive function of the spectrum of $R$ defined by initial conditions. For $n=r<l$, we have

\begin{equation}
 \frac{1}{l}(N - R)|r\rangle = 0|r\rangle,
\end{equation}
\noindent with $A_{0} = \alpha_{0}(R)$ such that $\alpha_{0}(R)|r\rangle=\alpha_{0}(r)|r\rangle$ 
and $R|r\rangle=r|r\rangle$. 

Now we will use this realization to show how to construct the Hamiltonian (\ref{Hamulthom})  from the transfer 
matrix (\ref{tutrihom}) and present their exact Bethe ansatz solution. It is straightforward to check that the Hamiltonian (\ref{Hamulthom}) is related with the transfer matrix $t(0)$ (\ref{t0trihom}), or with the conserved quantity $\mathcal{C}_0$ (\ref{cqmhn1}), through 

\begin{equation}
 H = \Omega \; t(0),
\end{equation}
\noindent where we have the following identification

\begin{eqnarray}
 \eta &=& \frac{l^2 U_{a} - lU_{ab} + U_{b}}{\Omega},
\end{eqnarray}

\begin{eqnarray}
 \theta &=& 2 l U_{ab} - 4U_{b}, \qquad \xi = 2 l^2 \mu_{a} - 2l\mu_{b},
\end{eqnarray}

\begin{eqnarray}
 2 l \Omega (\omega + \delta) &=& (2\Omega\eta  + \theta)N - \Omega\rho\eta + \xi,
\end{eqnarray}

\begin{eqnarray}
4l^2\Omega\omega\delta\eta + 8l^2\Omega\omega &=& \eta^2[\Omega\rho^{2}\eta + 4U_{b}N^2 - \theta N\rho \nonumber \\ 
&+& 4l(\Omega\omega + \mu_{b})N + (2l\Omega\omega - \xi)\rho], 
\end{eqnarray}
\noindent with $\rho\equiv\rho(R) = l\alpha_{0}(R)-2R$.

It is easy to see that $\rho$ is a conserved quantity using the total number of atoms, $N$, to write the conserved quantity $\mathcal{C}_1$ (\ref{cqmhn2}) as,

\begin{equation}
\mathcal{C}_1 = \frac{1}{l}N + \frac{1}{2l}\rho \label{cqmhn2b}.
\end{equation}

We can apply the algebraic Bethe ansatz method, using as the pseudo-vacuum 
the product state $(|0\rangle = |0 \rangle_b \otimes |r \rangle_A $, with $|0\rangle_b$ denoting the Fock vacuum state and $|r\rangle_A$ denoting the lowest weight state of 
the $sl(2)$ Lie algebra, where $r = 0,\;1,\;...\;,\; l-1,$ are the eigenvalues of $R$ for $N=nl+r$, with $n \in \mathbb{N}$,  
to find the Bethe ansatz equations (BAE)

\begin{equation}
\frac{(1-\eta v_{i}+\eta\delta)(v_{i}+\omega+\frac{\eta}{2}\alpha_{0}(r))}{v_{i}+\omega-\frac{\eta}{2}\alpha_{0}(r)}=\prod_{i\ne j}^{M}\frac{v_{i}-v_{j}-\eta}{v_{i}-v_{j}+\eta},\;\;\; i,j=1,...,M,
\label{BAEtrihomo}
\end{equation}
\noindent and the eigenvalues of the Hamiltonian (\ref{Hamulthom}),

\begin{eqnarray}
E & = & \Omega\eta^{-1}(\delta+\eta^{-1})\left(\omega+\frac{\eta}{2}\alpha_{0}(r)\right)\prod_{i=1}^{M}\frac{v_{i}-\eta}{v_{i}} \nonumber \\ &+& \Omega\eta^{-2}\left(\omega-\frac{\eta}{2}\alpha_{0}(r)\right)\prod_{i=1}^{M}\frac{v_{i}+\eta}{v_{i}}.
\label{energyhomo}
\end{eqnarray}

The parameters $\delta$ and $\omega$ are arbitrary and can be chosen conveniently. In the limit without scatterings, $U_j \rightarrow 0$, the BAE  (\ref{BAEtrihomo}) can be write as,

\begin{equation}
\sum_{i=1}^{M} \frac{1}{v_i + \omega} = \frac{1}{\alpha_0(r)}\sum_{i=1}^{M} v_i - \frac{M}{\alpha_0(r)}\delta,
\end{equation}
\noindent and for $\omega = 0$ the eigenvalues (\ref{energyhomo}) become,

\begin{equation}
E = \left(\frac{1}{2}\alpha_0(r) + M\right)\Omega\delta - \Omega\sum_{i=1}^{M} v_i.
\end{equation}
\noindent Now, the relation between the interconversion parameter and the externals potentials is simply,

\begin{equation}
\Omega = \frac{l \mu_{a} - \mu_{b}}{\delta}.
\end{equation} 

%
%

\section{Multiatomic hetero-nuclear molecular models}
     
In this section we present the integrability of a new family of Hamiltonians describing  multiatomic hetero-nuclear 
molecular BEC. The Hamiltonians that describes the interconversion of heterogeneous molecules labelled 
by $c$ with $l$ atoms of type $a$ and one atom of type $b$ are given by

\begin{eqnarray}
H &=& U_{a}N_a^2 + U_{b}N_b^2 + U_{c}N_c^2 + U_{ab}N_aN_b + U_{ac}N_aN_c + U_{bc}N_bN_c \nonumber \\ 
&+&  \mu_aN_a + \mu_bN_b + \mu_cN_c  \nonumber \\ 
&+& \Omega((a^{\dag})^l \; b^{\dag} \; c \; \alpha_{-}(N_a) + \alpha_{-}(N_a) \; c^{\dag} \; b \; (a)^l ),
\label{Hamulthet}
\end{eqnarray}
\noindent where $\alpha_{-}(N_a)$ is a function of $N_a$ that controls the amplitude of 
interconversion $\Omega$. In the same way of the Hamiltonians (\ref{Hamulthom}), this indicates that the density of atoms $N_a$ has some influence in the
generation of a bound-state composed by $l$ identical atoms. 

The imbalance between the number of atoms $a$ and the number of atoms $b$,

\begin{eqnarray}
\mathcal{J}_{ab} & = & N_{a}- l N_{b}, 
\label{cqmhet1}
\end{eqnarray}
\noindent is a conserved quantity and the total number of atoms, $ N = N_a + N_b + (l+1)N_c$, can be writes with the other two conserved quantities
\begin{eqnarray}
\mathcal{I}_{ac} & = & N_{a} + l N_{c}, \label{cqmhet2a} \\
 \mathcal{I}_{bc} & = & N_{b} +  N_{c}. \label{cqmhet2b}
\end{eqnarray}
 
Using $\mathcal{J}_{ab}$ and $N$, the $S$-wave diagonal part of the Hamiltonian (\ref{Hamulthet}) can be writte as,
\begin{equation}
 \alpha \mathcal{J}_{ab}^2 + \beta N^2 + \gamma N\mathcal{J}_{ab},
\end{equation} 
\noindent where we have used the following identification for the coupling constants

\begin{eqnarray}
U_{a} &=& \alpha + \beta + \gamma,  \qquad U_{b} = \alpha l^2 + \beta - \gamma l, \qquad U_{c} = \beta (l+1)^2,
\end{eqnarray}

\begin{eqnarray}
U_{ab} &=& -2l\alpha + 2\beta - \gamma (l-1), \qquad U_{ac} = 2\beta(l+1) + \gamma (l+1), 
\end{eqnarray}

\begin{eqnarray}
\;\;\;\;\;\;\;\;\;\;\;\;\;\;\;\;\;\;\;\;\;\; U_{bc} = 2\beta(l+1) - \gamma l(l+1).
\end{eqnarray}

Now, using the following realization for the $su(2)$ Lie algebra,

\begin{eqnarray}
S^{+} &=& b^{\dagger}c,\;\;\; S^{-}=c^{\dagger}b,\;\;\; S^{z}=\frac{N_{b}-N_{c}}{2},
\end{eqnarray}
\noindent and the multibosonic realization of the $sl(2)$ Lie algebra (\ref{multsl2}),

\begin{eqnarray}
A_{0} &=& \alpha_{0}(N), \;\;\; A_{-} = \alpha_{-}(N)a^l, \;\;\; A_{+} = (a^{\dagger})^l\alpha_{-}(N), 
\end{eqnarray}
\noindent with
\begin{eqnarray}
 \alpha_{0}(N) &=& \frac{2}{l}(N - R) + \alpha_{0}(R),  \\
 \alpha_{-}(N) &=& \sqrt{\frac{N!}{(N + l)!}(\frac{1}{l}(N - R) + \alpha_{0}(R))(\frac{1}{l}(N - R) + 1)},  
\end{eqnarray}
\noindent where $N = a^{\dagger}a$ and $l \in \mathbb{N}$, it is straightforward to check that the Hamiltonian (\ref{Hamulthet}) is related with the transfer matrix $t(u)$ (\ref{tutrihet}), or with the conserved quantity $\mathcal{C}_0$ (\ref{cqmhtn1}) if $u=0$, through

\begin{equation}
H = \sigma + \alpha \mathcal{J}_{ab}^2 + \beta N^2 + \gamma N \mathcal{J}_{ab}  + t(u),\end{equation} 

\noindent where the following identification has been made for the parameters

\begin{eqnarray}
\mu_{a} &=& 2\frac{u^{-}}{l}, \qquad \mu_{c}=-\mu_{b}=u^{+}, \qquad \Omega= \eta, \qquad \sigma = -\frac{u^{-}}{l}\rho,
\end{eqnarray}
\noindent with $\rho\equiv\rho(R) = l\alpha_{0}(R)-2R$.

We also can use the conserved quantities $\mathcal{J}_{ab}$ and $\mathcal{I}_{ac}$ to write the conserved quantity $\mathcal{C}_1$   (\ref{cqmhtn2}) as,

\begin{equation}
\mathcal{C}_1 = \frac{1}{l}(\mathcal{J}_{ab} + \mathcal{I}_{ac}) + \frac{1}{l}\rho,   \label{cqmhtn2a}
\end{equation}
\noindent showing that $\rho$ is also a conserved quantity.

We can apply the algebraic Bethe ansatz method, using as the pseudo-vacuum 
the product state $(|0\rangle = |r \rangle_A \otimes |\phi \rangle$,
with $|r\rangle_A$ denoting the lowest weight state of 
the $sl(2)$ Lie algebra where $r = 0,\;1,\;...\;,\; l-1,$ are the eigenvalues of $R$ for $N=nl+r$, with $n \in \mathbb{N}$ and $|\phi \rangle$ denoting the highest weight state of the $su(2)$ Lie algebra with weight $m_z$),
to find the Bethe ansatz equations (BAE)

\begin{equation}
-\frac{(v_{i} - \omega - \eta m_z)(v_{i} + \omega + \frac{\eta}{2} \alpha_0(r))}{(v_{i} - \omega + \eta m_z)(v_{i} + \omega - \frac{\eta}{2} \alpha_0(r))}=\prod_{i\ne j}^{M}\frac{v_{i}-v_{j}-\eta}{v_{i}-v_{j}+\eta},\;\;\; i,j=1,...,M,
\end{equation}
\noindent and the eigenvalues of the Hamiltonian (\ref{Hamulthet})

\begin{eqnarray}
E & = & \sigma + \alpha \mathcal{J}_{ab}^2 + \beta N^2 + \gamma N \mathcal{J}_{ab}  \nonumber \\ &+& (u - \omega - \eta m_z)(u + \omega + \frac{\eta}{2} \alpha_0(r))\prod_{i=1}^{M}\frac{u-v_{i}+\eta}{u-v_{i}} \nonumber \\ &-&(u - \omega + \eta m_z)(u + \omega - \frac{\eta}{2} \alpha_0(r))\prod_{i=1}^{M}\frac{u-v_{i}-\eta}{u-v_{i}}.
\label{eigenhet}
\end{eqnarray}
 
The eigenvalues (\ref{eigenhet}) are independent of the spectral parameter $u$ and of the parameter $\omega$, that are arbitrary.

\section{Summary}
I have introduced two new family of multiatomic  molecular BEC models for homo-nuclear and hetero-nuclear molecules and derived the Bethe ansatz equations  and the eigenvalues. The conserved quantities are also derived. The multiatomic homo-nuclear and hetero-nuclear molecular BEC models were obtained through a combination of Lax operators constructed using special realizations of the  $su(2)$ Lie algebra and Heisenberg-Weyl Lie algebra, as well as a multibosonic representation of the $sl(2)$ Lie algebra. The dependence of the parameters with the size of the molecules is explicit.

\subsection*{Acknowledgments}

The author acknowledge support from CNPq (Conselho Nacional de Desenvolvimento
Cient\'{\i}fico e Tecnol\'{o}gico). The author also would like to thank A. Foerster and I. Roditi for interesting discussions.


\section*{References}
\begin{thebibliography}{10}


\bibitem{early} E. A. Cornell and C. E. Wieman, {\it Rev. Mod. Phys.} \textbf{74} (2002) 875.

\bibitem{angly} J. R. Anglin and W. Ketterle, {\it Nature} \textbf{416} (2002) 211.

\bibitem{mol} P. Zoller, {\it Nature} \textbf{417} (2002) 493.

\bibitem{Feshbach} S. Inouye, M. R. Andrews, J. Stenger, H.-J. Miesner, D. M. Stamper-Kurn  and  W. Ketterle, {\it Nature} \textbf{392} (1998) 151.

\bibitem{RevFeshbach1} T. Kohler, K. Goral, and P. S. Julienne, {\it Rev. Mod. Phys.} \textbf{78} (2006) 1311.

\bibitem{RevFeshbach2} Cheng Chin, Rudolf Grimm, Paul Julienne and Eite Tiesinga, {\it Rev. Mod. Phys.} \textbf{82}, (2010) 1225.

\bibitem{RevPhotoassociation1} K. M. Jones, E. Tiesinga, P. D. Lett, and P. S. Julienne, {\it Rev. Mod. Phys.} \textbf{78} (2006) 483.

\bibitem{RevPhoto2} C. R. Menegatti, B. S. Marangoni and L. G. Marcassa, {\it Laser Physics} {\bf 18} (2008) 1305.

\bibitem{HomoCs} J. Herbig et al, {\it Science} \textbf{301} (2003) 1510.

\bibitem{HomoRb} S. Durr et al, {\it Phys. Rev. Lett.} \textbf{92} (2004) 020406.

\bibitem{HomoFeshbachCs2} C. Chin, T. Kraemer, M. Mark, J. Herbig, P. Waldburger, H.-C. N\"agerl and R. Grimm, {\it Phys. Rev. Lett.} \textbf{94} (2005) 123201.

\bibitem{HomoFeshbachDonley} E. A. Donley, N. R. Claussen, S. T. Thompson, and C. E. Wieman, {\it Nature} \textbf{417}, (2002) 529.

\bibitem{HetphotoKRb1} B. Damski, L. Santos, E. Tiemann, M. Lewenstein, S. Kotochigova, \\ P. Julienne, and P. Zoller, {\it Phys. Rev. Lett.} \textbf{90} (2003) 110401.

\bibitem{HetPhotoKRb2} M.W. Mancini, G. D. Telles, A. R. L. Caires, V. S. Bagnato and L. G. Marcassa, {\it Phys. Rev. Lett.}  \textbf{92} (2004) 133203.

\bibitem{HetPhotoNaCs} C. Haimberger, J. Kleinert, M. Bhattacharya and N. P. Bigelow, {\it Phys. Rev. A}  \textbf{70} (2004) 021402(R).

\bibitem{HetPhotoRbCs} J. M. Sage, S. Sainis, T. Bergeman, and D. DeMille, {\it Phys. Rev. Lett.} \textbf{94} (2005) 203001.

\bibitem{HetFeshbachRbRb} S. B. Papp and C. E. Wieman, {\it Phys. Rev. Lett.} \textbf{97} (2006) 180404.

\bibitem{HetFeshbachKRb} C. Weber, G. Barontini, J. Catani, G. Thalhammer, M. Inguscio and F. Minardi, {\it Phys. Rev. A} \textbf{78} (2008) 061601R.

\bibitem{HetFeshbachRbCs} K. Pilch, A. D. Lange, A. Prantner, G. Kerner, F. Ferlaino, H.-C. N\"agerl and R. Grimm, {\it Phys. Rev. A} \textbf{79} (2009) 042718.

\bibitem{HammerTeo} E. Braaten, H.-W. Hammer, {\it Annals of Physics} \textbf{322} (2007) 120.

\bibitem{EsryEfimovCs} B. D. Esry and C. Greene, {\it Nature} \textbf{440} (2006) 289.

\bibitem{EfimovEvidenceCs} T. Kraemer, M. Mark, P. Waldburger, J. G. Danzl, C. Chin, B. Engeser, 
A. D. Lange, K. Pilch, A. Jaakkola, H.-C. Nagerl, R. Grimm,
{\it Nature} \textbf{440} (2006) 315.

\bibitem{Homo3BEfimovCs2} S. Knoop, F. Ferlaino, M. Berninger, M. Mark, H.-C. N\"agerl and R. Grimm, {\it ̈Journal of Physics: Conf. Ser.} \textbf{194} (2009) 012064.

\bibitem{Homo4BEfimovCs} F. Ferlaino, S. Knoop, M. Berninger, W. Harm, J. P. D'Incao, H.-C. N\"agerl and R. Grimm, Phys. Rev. Lett. {\bf 102}   (2009) 140401.

\bibitem{HetEfimovKRb} G. Barontini, C. Weber, F. Rabatti, J. Catani, G. Thalhammer, M. Inguscio and F. Minardi, {\it Phys. Rev. Lett.} \textbf{103} (2009) 043201.

\bibitem{jon1} H.-Q. Zhou, J. Links, M. Gould and R. McKenzie,
{\it J. Math. Phys.} \textbf{44} (2003) 4690.

\bibitem{jon2} H.-Q. Zhou, J. Links, R. H. McKenzie, {\it Int. Jour. Mod. Phys. B} \textbf{17} (2003) 5819.

\bibitem{jonjpa} 
J. Links, H.-Q. Zhou, R. H. McKenzie and M. D. Gould, 
{\it J. Phys. A} \textbf{36} (2003) R63.

\bibitem{key-3} A. Foerster, J. Links, H.-Q. Zhou, in {\it Classical
and quantum nonlinear integrable systems: theory and applications},
edited by A. Kundu (IOP Publishing, Bristol and Philadelphia,
2003) pp. 208-233.

\bibitem{dukelskyy} J. Dukelsky, G. Dussel, C. Esebbag and S. Pittel,
{\it Phys. Rev. Lett.} \textbf{93} (2004) 050403.

\bibitem{Ortiz} G. Ortiz, R. Somma, J. Dukelsky and S. Rombouls, {\it Nuclear Physics B} \textbf{707} (2005) 421.
 
\bibitem{Kundu} A. Kundu, {\it Theoretical and Mathematical Physics} \textbf{151} (2007) 831.

\bibitem{eric5} A. Foerster and E. Ragoucy, {\it Nuclear Physics B} \textbf{777} (2007) 373.

\bibitem{GSantosaa} Jon Links, Angela Foerster, Arlei Prestes Tonel and Gilberto Santos, {\it Ann. Henri Poincar\'e}  \;{\bf 7} (2006) 1591.

\bibitem{GSantos} G. Santos, A. Foerster, I. Roditi, Z. V. T. Santos and A. P. Tonel, {\it J. Phys. A: Math. Theor.} \textbf{41} (2008) 295003.


\bibitem{GSantos1} G. Santos, A. Tonel, A. Foerster and J. Links, {\it Phys. Rev. A} \textbf{73} (2006) 023609.

\bibitem{GSantos2} G. Santos, A. Foerster, J. Links, E. Mattei and S. R. Dahmen, {\it Phys. Rev. A}  \textbf{73} (2010) 023609.

\bibitem{GSantos3} A. P. Tonel, C. C. N. Kuhn, G. Santos, A. Foerster, I. Roditi and Z. V. T. Santos, {\it Phys. Rev. A} \textbf{79} (2009) 013624.

\bibitem{Zhou1} L. Zhou, W. Zhang, H. Y. Ling, L. Jiang, and H. Pu, {\it Phys. Rev. A} \textbf{75} (2007) 043603.

\bibitem{Zhou2} L. Zhou, J. Qian, H. Pu, W. Zhang, and H. Y. Ling, {\it Phys. Rev. A} \textbf{78} (2008) 053612.

\bibitem{Vardi} A. Vardi, V. A. Yurovsky and J. R. Anglin, {\it Phys. Rev. A} \textbf{64} (2001) 063611.

\bibitem{maj} M. Duncan, A. Foerster, J. Links, E. Mattei, N. Oelkers and A. Tonel, 
{\it Nuclear Physics B} \textbf{767} [FS] (2007) 227.


\bibitem{hertier} M. H\'eritier, {\it Nature} \textbf{414} (2001) 31.

\bibitem{batchelor} M. T. Batchelor, {\it Physics Today} \textbf{60} (2007) 36.

\bibitem{fst} L. D. Faddeev, E. K. Sklyanin and L. A. Takhtajan,  {\it
Theor. Math. Phys.} {\bf 40} (1979) 194.

\bibitem{ks} P. P. Kulish and E. K. Sklyanin,  {\it Lect. Notes Phys.}
{\bf 151} (1982) 61. 

\bibitem{takhtajan} L. A. Takhtajan,  {\it Lect. Notes Phys.} {\bf 370} (1990)
3. 
\bibitem{korepin} V. E. Korepin, N. M. Bogoliubov and A. G. Izergin, {\it
Quantum inverse scattering method and correlation functions} (Cambridge
University Press, Cambridge, 1993).

\bibitem{faddeev} L. D. Faddeev,  {\it Int. J. Mod. Phys. A } \textbf{10} (1995)
1845.

\bibitem{korepin1} A. G. Izergin and V. E. Korepin, {\it Lett. Math. Phys.} {\bf 6} (1982) 283.
\bibitem{korepin2} A. G. Izergin and V. E. Korepin, {\it Nuc. Phys. B} \textbf{205} (1982) 401.
\bibitem{ek} F. H. L. Essler  and V. E. Korepin (eds.) 
{\it Exactly solvable models of strongly correlated
electrons} (World Scientific, Singapore, 1994). 

\bibitem{ek2} Fabian H. L. Essler, Holger Frahm, Frank G\"ohmann, Andreas Kl\"umper  and Vladimir E. Korepin, {\it The one-dimensional Hubbard Model} (Cambridge University Press, Cambridge, 2005). 
\bibitem{blz} V. Bazhanov, S. Lukyanov and A. B. Zamolodchikov,  {\it
Commun. Math. Phys.} \textbf{177} (1996) 381.
\bibitem{jimbo85} M. Jimbo,  {\it Lett. Math. Phys.} \textbf{10} (1985) 63. 

\bibitem{jimbo86} M. Jimbo,  {\it Lect. Notes Phys.} \textbf{246} (1986) 335.

\bibitem{drinfeld} V. G. Drinfeld,  Quantum groups {\it Proc. Int. Congress of
Mathematicians} ed A M Gleason (Providence, RI: American Mathematical
Society) (1986) 798.

\bibitem{frt} N. Yu Reshetikhin,  L. A. Takhtajan and L. D. Faddeev,  {\it
Leningrad Math. J.} \textbf{1} (1990) 193.
\bibitem{faddeev2}  L. D. Faddeev, 40 Years in Mathematical Physics {\it World Scientific Series in 20th Century Mathematics}, vol. 2, (World Scientific Publishing Co. Pte. Ltd., Singapore, 1995).


\bibitem{Tomasz} T. Goli\'{n}ski, M. Horowski, A. Odzijewicz, A. Sli\.{z}ewska, {\it Jour. Math. Phys.} \textbf{48} (2007)  023508.

\bibitem{Roditi} I. Roditi, {\it Brazilian Journal of Physics} \textbf{30} (2000) 357.  



\end{thebibliography}
\end{document}